\documentclass[twoside]{article}
\usepackage[a4paper]{geometry}
\usepackage[latin1]{inputenc} 
\usepackage[T1]{fontenc} 
\usepackage{RR}
\usepackage{hyperref}

\usepackage{latexsym,amsfonts,amsmath}
\usepackage{graphicx}
\usepackage{epstopdf}

{\bfseries}{\itshape}

{\bfseries}{\itshape}

\newtheorem{theorem}{Theorem}[section]{\bfseries}{\itshape}

\newtheorem{corollary}{Corollary}[section]{\bfseries}{\itshape}

\newtheorem{proposition}{Proposition}[section]{\bfseries}{\itshape}

{\bfseries}{\itshape}

\newtheorem{lemma}{Lemma}[section]{\bfseries}{\itshape}

\newtheorem{remark}{Remark}[section]{\bfseries}{\itshape}

{\bfseries}{\itshape}

{\bfseries}{\itshape}

\RRdate{March 2015}
\RRauthor{
Konstantin Avrachenkov\thanks{Inria Sophia Antipolis, France, {\tt k.avrachenkov@sophia.inria.fr}}%
  \and
Alexey Piunovskiy\thanks{University of Liverpool, United Kingdom,
{\tt piunov@liv.ac.uk}}
\and \\
Yi Zhang\thanks{University of Liverpool, United Kingdom, {\tt yi.zhang@liv.ac.uk}}
}

\authorhead{K. Avrachenkov \& A. Piunovskiy \& Y. Zhang}
\RRtitle{Le Temps de Premier Passage dans les Processus de Markov avec Red\'emarrage
et leur Application \`a la Mesure de Centralit\'e de R\'eseau
}
\RRetitle{Hitting Times in Markov Chains with Restart\\
and their Application to Network Centrality}
\titlehead{Hitting Times in Markov Chains with Restart}
\RRresume{
Motiv\'e par diverses applications provenant de t\'el\'ecommunications, informatique
et de la physique, nous considerons un processus generale de Markov dans l'espace de Borel
avec une possibilit\'e de red\'emarrage. \`A chaque \'etape, avec une probabilit\'e le processus
red\'emarre a partir d'une distribution donn\'ee et avec la probabilit\'e compl\'ementaire
le processus continue l'\'evolution selon une noyau de Markov.
Nous \'etudions l'esp\'erance du temps de premier passage
\`a l'ensemble donn\'e. Nous obtenons une formule explicite pour l'esp\'erance du temps
de premier passage et d\'emontrons que le processus avec red\'emarrage est Harris positif r\'ecurrent.
Ensuite, nous \'etablissons que les assertions suivantes sont \'equivalentes :
(a) le fait d'\^etre limit\'e (par rapport \`a l'\'etat initiale) de l'esp\'erance du temps
de premier passage; (b) la finitude de l'esp\'erance du temps de premier passage
pour presque tous les \'etats initiaux par rapport \`a la probabilit\'e invariante unique;
et, (c) l'ensemble cible est de mesure positive par rapport \`a la
probabilit\'e invariante.  Enfin, nous illustrons nos r\'esultats th\'eoriques avec
deux exemples dans les espaces d\'enombrables et non d\'enombrables et avec l'application
\`a la mesure de centralit\'e de r\'eseau.
}

\RRabstract{
Motivated by applications in telecommunications, computer science
and physics, we consider a discrete-time Markov process with
restart. At each step the process either
with a positive probability restarts from a given distribution, or
with the complementary probability continues according to a Markov
transition kernel. The main contribution of the present work is that
we obtain an explicit expression for the expectation of the hitting
time (to a given target set) of the process with restart.
The formula is convenient when considering the problem of optimization
of the expected hitting time with respect to the restart probability.
We illustrate our results with two examples
in uncountable and countable state spaces and
with an application to network centrality.
}
\RRmotcle{Processus de Markov en Temps Discret, L'Esp\'erance du Temps
de Premier Passage, La Mesure de Centralit\'e de R\'eseau}

\RRkeyword{Discrete-time Markov Process with Restart, Expected Hitting Time,
Network Centrality}

\RRprojets{Neo}
\URSophia 

\begin{document}
\RRNo{8581}
\makeRR   

\section{Introduction}
We give a self-contained study of a discrete-time Markov process
with restart. At each step the process either with the positive
probability $p$ restarts from a given distribution, or with the
complementary probability $1-p$ continues according to a Markov
transition kernel. Such processes have many applications in
telecommunications, computer science and physics. Let us cite just
a few. Both TCP (Transmission Control Protocol) and HTTP
(Hypertext Transfer Protocol) can be viewed as protocols
restarting from time to time, \cite{KR01}, \cite{MH01}.
The PageRank network centrality \cite{BP98}, in information
retrieval, models the behaviour of an Internet user surfing the web
and restarting from a new topic from time to time.
The sybil attack resistant network centralities based on the hitting
times of a random walk with restart have been proposed in \cite{HS08,Liuetal16}.
Markov processes with restart are useful for the analysis of replace
and restart types protocols in computer reliability, \cite{AFLRS08},
\cite{ALT14}, \cite{KNT87}.
The restart policy is also used to speedup the Las Vegas type randomized
algorithms \cite{Alt96}, \cite{LSZ93}.
Finally, human and animal mobility patterns can be modeled by Markov processes
that restart from some locations \cite{GHB08}, \cite{WBC10}.

The main focus of the present work is the expectation of the hitting time (to a given target set) of the
process with restart, for which we obtain simple explicit expressions in terms of the expected discounted
hitting time of the original process without restart; see Theorem \ref{thm:explsol}.
These formula is useful in the optimization of the expected hitting time with respect to the restart probability. In addition,
the formulae allow us to either refine, or give simple and self-contained proofs of the stability results for the process with restart in terms of hitting times. Finally, in Section~\ref{APZ2014Section3},
we illustrate the main results with two examples in uncountable and countable state spaces and with an application to network centrality.
In particular, we show that the hitting time based network centrality can be more discerning than PageRank.

Let us mention some related work to the present one in the current
literature. The continuous-time Markov process with restart was
considered in \cite{APZ13}. According to Theorem 2.2 in \cite{APZ13},
the continuous-time Markov process with restart is
positive Harris recurrent in case the original process is honest.
At the same time, the process with restart is not positive Harris
recurrent if the original process is not honest (i.e., the
transition kernel is substochastic; in case the state space is
countable, that means the accumulation of jumps). The objective
of \cite{APZ13} does not lie in the expected hitting time,
but in the representation of the transition probability function of the
(continuous-time) process with restart in terms of the one of the
original (continuous-time) process without restart. This is
trivial in the present discrete-time setup. Here our focus is on
the characterization of the expected hitting time. We also would like
to mention the two works \cite{DTW03} and \cite{JP12}, dealing with the control
theoretic formulation, where the controller decides (dynamically)
whether it is beneficial or not to perform a restart at the
current state. That line of research can be considered
complementary to ours.

The rest of this paper is organized as follows. The description
of the process with restart and the main statements are presented in
Section \ref{APZ2014Sec2}, which are illustrated by two examples
and network centrality application in Section \ref{APZ2014Section3}.
The paper is finished with a conclusion in Section \ref{APZ2014Conclusion}.

\section{Main statements}\label{APZ2014Sec2}

Let us introduce the model formally. As in \cite{Meyn:1993}, let $E$ be a nonempty locally compact Borel state
space endowed with its Borel $\sigma$-algebra ${\cal B}(E).$
Consider a discrete-time Markov chain
$\tilde{X}=\{\tilde{X}_t,~t=0,1,\dots\}$ in the state space
$E$ with the transition probability function $\tilde{P}(x,dy)$
being defined by
\begin{eqnarray}\label{APZ2014jtp1}
\tilde{P}(x,\Gamma):=p \nu(\Gamma)+(1-p)P(x,\Gamma),
\end{eqnarray}
for each $\Gamma\in {\cal B}(E)$, where $p\in(0,1)$ and $P(x,dy)$
is a transition probability function, and $\nu$ is a probability
measure on ${\cal B}(E).$ Let $X:=\{X_t,t=0,1,\dots\}$ denote the
Markov chain corresponding to the transition probability
$P(x,dy)$. We assume the two processes $X$ and $\tilde{X}$ are
defined on the common probability space $(\Omega, \cal F,
\mathbb{P})$; if we emphasize that the initial state is $x\in E,$
then $E_x$ denotes the corresponding expectation operator, and the
notation $P_x$ is similarly understood.

The process $\tilde{X}$ is understood as the modified version of
the process $X$, and is obtained by restarting (independently of
anything else) the process $X$ after each transition with
probability $p\in(0,1)$ and the distribution of the state after
each restart being given by $\nu;$ whereas if there is no restart
after the transition (with probability $1-p$), the distribution of
the post-transition state is $P(x,dy)$ (given that the current
state is $x$).

The following notation is used throughout this paper. Let
$P^t(x,dy),~t=0,1,\dots,$ be defined iteratively as follows; for
each $\Gamma\in {\cal B}(E),$
\begin{eqnarray*}
P^0(x,\Gamma)&:=&I\{x\in \Gamma\},\\
P^{t+1}(x,\Gamma)&:=&\int_E P^{t}(x,dy)P(y,\Gamma).
\end{eqnarray*}
The power of other kernels is understood similarly.

Finally, thoughout this paper, the convention $0\cdot+\infty:=0$ is in use.

\subsection{Known facts}
The materials in this subsection are standard and known from \cite{Meyn:1993}. The purpose here is to give a short and self-contained presentation. The main result of this paper is postponed to the next subsection.

From (\ref{APZ2014jtp1}) it is clear that the process $\tilde{X}$ is Harris recurrent with an irreducibility measure $\nu$. (Recall that the process $\tilde{X}$ is $\nu$-irreducible if for each set
$\Gamma\in {\cal B}(E)$ satisfying $\nu(\Gamma)>0$, it holds that
$P_x(\tau_\Gamma<\infty)>0$ for each $x\in E,$ where
\begin{eqnarray}\label{APZ2014NNNewNNN}
\tau_\Gamma:= \inf\{t=1,2,\dots:\tilde{X}_t\in \Gamma\}.
\end{eqnarray} As usual, $\inf \emptyset:=\infty.$)
If a Harris recurrent process admits an invariant probability, then it is called positive Harris recurrent; in that case the invariant probability is unique. We verify that $\tilde{X}$ is positive Harris recurrent with the unique invariant probability $q$ given in the next statement.

\begin{proposition}\label{APZ2014Thm1}
The process $\tilde{X}$ is positive Harris recurrent with the
unique invariant probability measure $q(dy)$ given by
\begin{eqnarray}\label{APZ2014InvariantmeasureFormula}
q(\Gamma)=\int_E \sum_{t=0}^\infty p(1-p)^t P^t(y,\Gamma)\nu(dy)
\end{eqnarray}
for each $\Gamma\in {\cal B}(E).$
\end{proposition}

\par\noindent\textit{Proof.}
Clearly, $q(dy)$ is a probability measure, and routine calculations verify
\begin{eqnarray*}
q(\Gamma)=\int_E q(dx)\tilde{P}(x,\Gamma)
\end{eqnarray*}
for each $\Gamma\in {\cal B}(E)$.   $\hfill\Box$
\bigskip

We strengthen the above statement in the next corollary.
Observe that the process $\tilde{X}$ is aperiodic in the sense of p.118 of \cite{Meyn:1993}, and the state space $E$ is a petite set  since $\tilde{P}(x,\cdot)\ge p \nu(\cdot)$ for each $x\in E$. By Theorem 16.0.2 (vi) of \cite{Meyn:1993}, we see that the process $\tilde{X}$ is uniformly ergodic (see also related arguments in \cite{N02,NT82}). However, in the next statement, we give a direct simple proof of this fact, and obtain the rate of convergence in terms of the restart probability. Below, $||\cdot||_{TV}$ stands for the total variation norm of finite signed measures.
\begin{corollary}\label{APZ2014UniErg}
The process $\tilde{X}$ is uniformly ergodic with the
unique invariant probability measure $q(\cdot)$ given in Proposition \ref{APZ2014Thm1}.
In particular, we have
\begin{equation}
\label{UniErgIneq}
||\tilde P^n(x,\cdot)-q(\cdot)||_{TV} \le 2(1-p)^n.
\end{equation}
\end{corollary}
\par\noindent\textit{Proof.}
We first note that
\begin{eqnarray*}
\tilde P^n(x,\Gamma) = \int_E \sum_{t=0}^{n-1} p(1-p)^t P^t(y,\Gamma) \nu(dy)
+ (1-p)^n P^n(x,\Gamma),
\end{eqnarray*}
which can be easily shown by induction. Next, using the above expression and Proposition \ref{APZ2014Thm1}, we can write
\begin{eqnarray*}
&&||\tilde P^n(x,\cdot)-q(\cdot)||_{TV}
=
||-\int_E \sum_{t=n}^{\infty} p(1-p)^t P^t(y,\cdot) \nu(dy) + (1-p)^n P^n(x,\cdot)||_{TV}\\
&\le& (1-p)^n ||-\int_E \sum_{t=0}^{\infty} p(1-p)^t P^{t+n}(y,\cdot) \nu(dy) + P^n(x,\cdot)||_{TV}\\
&
\le& 2(1-p)^n.
\end{eqnarray*}
Thus, we have established inequality (\ref{UniErgIneq}) and hence the uniform ergodicity.
$\hfill\Box$

\subsection{Hitting times}

In this paper we are primarily interested in the expected hitting
time of the process $\tilde{X}$ to a given set $H\in {\cal B}(E),$
defined by
\begin{eqnarray*}
\eta_H:=\inf\{t=0,1,\dots:\tilde{X}_t\in H\}.
\end{eqnarray*}
For the future reference, we put
\begin{eqnarray*}
{}_H P^0(x,E):=I\{x\in E\setminus H\}.
\end{eqnarray*}
Denote
\begin{eqnarray}\label{APZ2014Defagain}
V(x) := E_x\left[\eta_H\right]
\end{eqnarray}
 and let
\begin{eqnarray*}
{}_H P(x,\Gamma) =P(x,\Gamma\setminus H)
\end{eqnarray*}
be the taboo transition kernel with respect to the set $H$. Then,
one can write
\begin{eqnarray}
\label{eq:dynprog} &&V(x) = 1 + p \int_E V(y) \nu(dy) + (1-p)
\int_E V(y) {}_H P(x,dy), \quad \forall~x\in E\setminus H,\nonumber\\
&& V(x) = 0, \quad \forall~ x \in H.
\end{eqnarray}
Furthermore, it is well known that the function $V(x)$ defined by
(\ref{APZ2014Defagain}) is the minimal nonnegative (measurable)
solution to equation (\ref{eq:dynprog}), and can be obtained by
iterations
\begin{eqnarray*}
V^{(n+1)}(x) = 1 + p \int_E V^{(n)}(y) \nu(dy)
+ (1-p) \int_E V^{(n)}(y) {}_H P(x,dy),~x\in E\setminus H, \quad n=0,1,...
\end{eqnarray*}
with $V^{(n)}(x) = 0$ if $x \in H$ for each $n=0,1,\dots$, and
$V^{(0)}(x) \equiv 0$; c.f. e.g., Proposition 9.10 of \cite{Bertsekas:1978}.

One can actually obtain the minimal nonnegative solution to
(\ref{eq:dynprog}) in the explicit form.
\begin{theorem}
\label{thm:explsol} \par\noindent(a) The minimal nonnegative
solution to (\ref{eq:dynprog}) is given by the following explicit
form
\begin{eqnarray}
\label{eq:explsol} &&V(x) = V_1(x)\sum_{t=0}^\infty
\left(p\int_{E} V_1(y)\nu(dy)\right)^t, \quad \forall~x\in E\setminus H,\nonumber\\
&&V(x)=0, \quad \forall~x\in H,
\end{eqnarray}
where the function $V_1$ is given by
\begin{eqnarray}
\label{eq:V1} &&V_1(x) := \sum_{t=0}^\infty (1-p)^t {}_H
P^t(x,E), \quad \forall~x\in E\setminus H;\nonumber\\
&&V_1(x):=0, \quad \forall~x\in H.
\end{eqnarray}
It coincides with the unique bounded solution to the equation
\begin{eqnarray}\label{APZ2014DPEqn2}
&&V_1(x)=1+(1-p)\int_E V_1(y) {}_H P(x,dy), \quad \forall~x\in E\setminus H;\nonumber\\
&& V_1(x)=0, \quad \forall~x\in H.
\end{eqnarray}
\par\noindent(b) If $q(H)>0,$ then
\begin{eqnarray}
\label{eq:explsol222} && V(x) = \frac{V_1(x)}{1 - p\int_{E} V_1(y)
\nu(dy)}<\infty, \quad \forall~x\in E\setminus H,\nonumber\\
&&V(x)=0, \quad \forall~x\in H.
\end{eqnarray}
\end{theorem}

\par\noindent\textit{Proof.} (a) Observe that the function $V_1$ given by (\ref{eq:V1})
represents the expected total discounted time before the first
hitting of the process $X$ at the set $H$ given the initial state
$x$ and the discount factor $1-p.$ It thus follows from the
standard result about the discounted dynamic programming with a
bounded reward that the function $V_1$ is the unique bounded
solution to equation (\ref{APZ2014DPEqn2}); see e.g., Theorem
8.3.6 of \cite{Hernandez-Lerma:1999}.

Now by multiplying both sides of the equation
(\ref{APZ2014DPEqn2}) by the expression
\begin{eqnarray*}
\sum_{t=0}^\infty \left(p\int_{E} V_1(y)\nu(dy)\right)^t
\end{eqnarray*}
for all $x\in E\setminus H,$ it can be directly
verified that the function $V$ defined in terms of $V_1$ by
(\ref{eq:explsol}) is a nonnegative solution to
(\ref{eq:dynprog}). We show that it is indeed the minimal
nonnegative solution to (\ref{eq:dynprog}) as follows.

Let $U \ge 0$ be an arbitrarily fixed nonnegative solution to
(\ref{eq:dynprog}). It will be shown by induction that
\begin{eqnarray}\label{APZPiunov1}
U(x)\ge V_1(x)\sum_{t=0}^n \left(p \int_{E}
V_1(y)\nu(dy)\right)^t~\forall~n=0,1,\dots,~\forall~x\in E.
\end{eqnarray}

The case when $x\in H$ is trivial.

Let $x\in E\setminus H$ be arbitrarily fixed. It follows from
(\ref{eq:dynprog}) that
\begin{eqnarray}\label{APZ2014ProofEqnEqn11}
U(x)\ge 1+(1-p)\int_E {}_HP(x,dy)=\sum_{t=0}^1(1-p)^t \left(\int_E
{}_HP^t(x,dy)\right) \ge 1.
\end{eqnarray}
If for some $n\ge 1$
 \begin{eqnarray}\label{APZ2014ProofEqnEqn1}
U(x)\ge
\sum_{t=0}^n(1-p)^t \left(\int_E {}_HP^t(x,dy)\right),
\end{eqnarray}
then by (\ref{eq:dynprog}),
\begin{eqnarray*}
U(x)&\ge& 1+(1-p)\int_E  U(y){}_HP(x,dy)\\
&\ge & 1+(1-p)\int_E  \left(\sum_{t=0}^n(1-p)^t \left(\int_E
{}_HP^t(y,dz)\right)\right){}_HP(x,dy)\\
&=& \sum_{t=0}^{n+1}(1-p)^t \left(\int_E {}_HP^t(x,dy)\right),
\end{eqnarray*}
and so (\ref{APZ2014ProofEqnEqn1}) holds for all $n\ge 0$ and thus
by (\ref{eq:V1})
\begin{eqnarray*}
U(x)\ge V_1(x).
\end{eqnarray*}
Consequently, (\ref{APZPiunov1}) holds when $n=0.$

Suppose (\ref{APZPiunov1}) holds for $n$, and consider the case of
$n+1.$ Then from (\ref{eq:dynprog}),
\begin{eqnarray}\label{APZ2014ProofOKay2}
U(x) &=& 1+p \int_E U(y)\nu(dy)+(1-p)\int_E  U(y){}_HP(x,dy)\nonumber\\
&\ge& \sum_{t=0}^{n+1} \left(p\int_E
V_1(y)\nu(dy)\right)^t+(1-p)\int_E  U(y){}_HP(x,dy),
\end{eqnarray}
where the inequality follows from the inductive supposition.
Define the function $W$ on $E$ by
\begin{eqnarray*}
W(z)=\frac{U(z)}{\sum_{t=0}^{n+1} \left(p\int_E
V_1(y)\nu(dy)\right)^t}, \quad \forall~z\in E.
\end{eqnarray*}
Then by (\ref{APZ2014ProofOKay2}),
\begin{eqnarray*}
W(x)\ge 1+(1-p)\int_E  {}_HP(x,dy)=\sum_{t=0}^1(1-p)^t
\left(\int_E {}_HP^t(x,dy)\right);
\end{eqnarray*}
c.f. (\ref{APZ2014ProofEqnEqn11}). Now, based on
(\ref{APZ2014ProofOKay2}), a similar reasoning by induction as to
the verification of (\ref{APZ2014ProofEqnEqn1}) for all $n\ge 0$
shows that
\begin{eqnarray*}
W(x)\ge \sum_{t=0}^k(1-p)^t \left(\int_E {}_HP^t(x,dy)\right),
\quad \forall~k=0,1,\dots,
\end{eqnarray*}
and thus $W(x)\ge V_1(x)$. This means
\begin{eqnarray*}
U(x)\ge V_1(x)\sum_{t=0}^{n+1} \left(p\int_E
V_1(y)\nu(dy)\right)^t.
\end{eqnarray*}
Thus by induction, (\ref{APZPiunov1}) holds for all $n\ge 0$, and
thus
\begin{eqnarray*}
U(x)\ge V(x)
\end{eqnarray*}
by (\ref{eq:explsol}), as desired.

(b) If $q(H)>0$, then there exists some $T\ge 0$ such that $\int_E
P^T(y,H)\nu(dy)>0$, meaning that there exists some $T'\le T$ such
that $\int_E {}_HP^{T'}(x,E)\nu(dx)<1$. Thus,
\begin{eqnarray*}0\le
p\int_ E\sum_{t=0}^\infty (1-p)^t {}_H P^t(y,E)\nu(dy)<1,
\end{eqnarray*}
and so the geometric series $\sum_{t=0}^\infty \left(p\int_{E}
V_1(y)\nu(dy)\right)^t$ converges. The statement follows.
$\hfill\Box$

\bigskip

The next corollary is immediate.
\begin{corollary}\label{APZ2014V1bound}
If $q(H)>0$, then both $V_1(x)$ and $V(x)$ are bounded with respect to the state
$x\in E$. In particular, we have
\begin{eqnarray*}
V_1(x) \le \frac{1}{p}.
\end{eqnarray*}
\end{corollary}

There is a nice probabilistic interpretation of the decomposition presented in Theorem~\ref{thm:explsol}.
Suppose $q(H)>0,$ which is case~(b) in Theorem~\ref{thm:explsol}.
Let us make a change of variable
\begin{eqnarray*}
\rho(x) = 1 - p V_1(x)
= 1 - p \sum_{t=0}^\infty (1-p)^t {}_H P^t(x,E) = P_x[\mbox{hit before restart}].
\end{eqnarray*}
Then, equation (\ref{eq:explsol222}) takes the form
$$
V(x) = \frac{1-\rho(x)}{p\int_E \rho(y) \nu(dy)}
= \frac{P_x[\mbox{no hit before restart}]}{p P_\nu[\mbox{hit before restart}]} .
$$
The expected number of steps in a cycle from one restart to another is $1/p$, and the expected
number of cycles until we hit the set $H$ is roughly $1/P_\nu[\mbox{hit before restart}]$.
Thus, the total expected steps is approximately equal to $1/(pP_\nu[\mbox{hit before restart}])$.
This is only an approximation since we did not carefully take into account the impact of the first
and the last incomplete cycles.
It turns out that the factor $P_x[\mbox{no hit before restart}]$ gives the necessary correction.
Theorem~\ref{thm:explsol} can be viewed as a generalization of the results of \cite{AL06,HS08}
to the general state space from the finite state space.

We also have the following chain of equivalent statements.
\begin{theorem}\label{APZ2014Equivalencetheorem}
The following statements are equivalent.
\par\noindent(a) $q(H)>0.$
\par\noindent(b) $V(x)=E_x[\eta_H]<\infty$ for each $x\in E.$
\par\noindent(c) $V(x)=E_x[\eta_H]<\infty$ for almost all $x\in E$ with
respect to $q(dy).$
\par\noindent(d) $\sup_{x\in E}V(x)=\sup_{x\in E}E_x[\eta_H]<\infty$.
\end{theorem}
The proof of this theorem follows from the next lemma.
\begin{lemma}\label{APZ2014Thm2}
Statements (a) and (c) in Theorem \ref{APZ2014Equivalencetheorem}
are equivalent.
\end{lemma}
\par\noindent\textit{Proof.}
Statement (a) implies statement (c) by Theorem
\ref{thm:explsol}(b). It remains to show that statement (c)
implies statement (a). To this end, suppose for contradiction that
statement (a) does not hold, i.e., $q(H)=0.$ Then it follows from
(\ref{APZ2014InvariantmeasureFormula}) that
\begin{eqnarray*}
\int_E \sum_{t=0}^\infty P^t(y,H)\nu(dy)=0.
\end{eqnarray*}
Thus, there exists a measurable subset $\Gamma$ of $E\setminus H$
such that
\begin{eqnarray}\label{APZ2014G1}
\nu(\Gamma)=1
\end{eqnarray}
and \begin{eqnarray}\label{APZ2014G2}
P_x[\eta_H^X=\infty]=1~\forall~x\in \Gamma,
\end{eqnarray}
where
\begin{eqnarray}\label{APZ2014NewNotation}
\eta_H^X:=\inf\{t=0,1,\dots:X_t\in H\}.
\end{eqnarray}
Let $x\in \Gamma$ be fixed. Then by (\ref{eq:V1}) and
(\ref{APZ2014G2})
\begin{eqnarray*}
V_1(x)= E_x\left[\sum_{t=0}^{\eta^X_H-1}
(1-p)^t\right]=E_x\left[\sum_{t=0}^\infty
(1-p)^t\right]=\frac{1}{p}.
\end{eqnarray*}
Since $x\in \Gamma$ is arbitrarily fixed, and $\nu(\Gamma)=1$ by
(\ref{APZ2014G1}), we see from (\ref{eq:explsol}) and the fact
that $V_1(y)\ge 1$ if $y\in E\setminus H$ that $V(y)=\infty$ for
each $y\in E\setminus H$. Since $q(E\setminus H)=1,$ we see that
statement (c) does not hold. $\hfill\Box$
\bigskip

\par\noindent\textit{Proof of Theorem
\ref{APZ2014Equivalencetheorem}.} From Theorem
\ref{thm:explsol}(b), we see that (a) implies (b), which implies
(c). From Lemma \ref{APZ2014Thm2}, (c) implies (a). Clearly, (d)
implies (b). Finally, (a) implies (d) because $V_1(x)$ is bounded;
see Corollary \ref{APZ2014V1bound}. $\hfill\Box$

\begin{remark}
Corollary \ref{APZ2014UniErg} and Theorem 16.0.2 of \cite{Meyn:1993} immediately give the relation $(a)\Rightarrow (d)\Rightarrow (b)\Rightarrow (c)$ in Theorem
\ref{APZ2014Equivalencetheorem}. However, we gave its self-contained proof in the above without referring to the result concerning the general Markov chains in \cite{Meyn:1993}.
\end{remark}

\subsection{Optimization problem}

Next let us consider the dependance of $V(x)$ on the restart
probability $p$ for each fixed $x\in E$. When we emphasize the
dependance on $p$, we explicitly write $V(x,p)$ and $V_1(x,p).$ In
the above, $p\in(0,1)$ was fixed. Now we formally put
\begin{eqnarray*}
&&V(x,0):=\sum_{t=0}^\infty {}_H
P^t(x,E)\in[1,\infty], \forall~x\in E\setminus H;\\
&&V(x,0):=0, \forall~x\in H,
\end{eqnarray*}
which represents the expected hitting time of the process without
restart, and
\begin{eqnarray*}
&&V(x,1):=\frac{1}{\nu(H)}\in[1,\infty], \forall~x\in E\setminus H;\\
&&V(x,1):=0, \forall~x\in H,
\end{eqnarray*}
which represents the expected hitting time of the process that
restarts with full probability at each transition. Here and below,
$\frac{c}{0}:=\infty$ for any $c>0.$ As usual, the continuity of
$V(x,p)$ at $p=a$ means $\lim_{p\rightarrow a}
V(x,p)=V(x,a)\in[-\infty,\infty].$

\begin{theorem}\label{APZ2014Continiutytheorem}
Suppose \begin{eqnarray}\label{APZ2014NewCondition}
 \int_E \sum_{t=0}^\infty
P^t(y,H)\nu(dy)>0.
\end{eqnarray} Then the function $V(x,p)$ is infinitely many times differentiable
in $p \in (0,1)$ and is continuous in $p\in[0,1].$ As a
consequence, the problem
\begin{eqnarray}\label{APZ2014P1}
\mbox{Miniminize $ V(x,p)$ with respect to $p\in[0,1]$ }
\end{eqnarray}
is solvable.
\end{theorem}
\begin{remark}
The condition (\ref{APZ2014NewCondition}) in the above statement
is equivalent to $q(H)>0$ for some and then all $p\in(0,1)$ by
(\ref{APZ2014InvariantmeasureFormula}).
\end{remark}

\par\noindent\textit{Proof of Theorem \ref{APZ2014Continiutytheorem}.} If $x\in H,$ the statement holds
trivially since $V(x,p)=0$ for each $p\in[0,1].$ Consider now
$x\in E\setminus H.$
One can see that $V_1(x,p)$ (as given by (\ref{eq:V1})) and $(1 -
p\int_{E} V_1(y,p) \nu(dy))^{-1}$ are both infinitely many times
differentiable in $p\in(0,1).$ It follows from this and
(\ref{eq:explsol222}) that $V(x,p)$ is infinitely many times
differentiable in $p\in(0,1).$ For the the continuity of $V(x,p)$
at $p=0,$ it holds that
$$
1-p\int_E V_1(y,p)\nu(dy) = 1- p\int_E
E_y\left[\sum_{t=0}^{\eta^X_H-1} (1-p)^t\right]\nu(dy)
$$
$$
= 1-\int_E E_y\left[1-(1-p)^{\eta^X_H}\right]\nu(dy)
= \int_E E_y\left[(1-p)^{\eta^X_H}\right]\nu(dy)\rightarrow 1,
$$
as $p\rightarrow 0$, by the monotone convergence theorem; recall
(\ref{APZ2014NewNotation}) for the definition of $\eta_H^X$. It
follows from this fact, (\ref{eq:V1}), (\ref{eq:explsol222}) and
the monotone convergence theorem that
\begin{eqnarray*}
\lim_{p\rightarrow 0} V(x,p) &=& \frac{\lim_{p\rightarrow
0}V_1(x,p)}{\lim_{p\rightarrow 0}\left\{1 - p\int_{E} V_1(y,p)
\nu(dy)\right\}}= V(x,0)
\end{eqnarray*}
as desired.  The continuity of the function $V(x,p)$ at $p=1$ can
be similarly established. The last assertion is a well known fact;
see e.g., \cite{Bertsekas:1978}. $\hfill\Box$

\bigskip

We shall illustrate the optimization problem (\ref{APZ2014P1})
by two examples in the next section.


\section{Numerical examples and application}\label{APZ2014Section3}

\subsection{Uni-directional random walk on the line}\label{APZ2014ExampleExponential}
Let $E=\mathbb{R}$, $H=[a,b]$ with $a<b$ being two real numbers.
The process $X$ only moves to the right, and the increments of
each of the transitions are i.i.d. exponential random variables
with the common mean $\frac{1}{\mu}>0$. The restart probability is
denoted as $p\in(0,1)$ as usual, and the restart distribution
$\nu$ is arbitrary. Below by using Theorem \ref{thm:explsol} we
provide the explicit formula for the expected hitting time at $H$
of the restarted process $\tilde{X}$. (Clearly, if the initial
state is outside $H$, then the expected hitting time of the
process $X$ at the set $H$ is infinite.)

We can give the following informal description of this example.
There is a treasure hidden in the interval $[a,b]$ and one tries
to find the treasure. Once the searcher checks one point in the
interval $[a,b]$, he finds the treasure. The searcher has the
means only to stop and to check points between the exponentially
distributed steps. This models the cost of checking frequently. It
is also natural to restart the search from some base. Intuitively,
by restarting too frequently, the searcher spends most of the time
near the base and does not explore the area sufficiently. On the
other hand, restarting too seldom leads the search to very far
locations where the searcher spends most of the time for nothing.
Hence, intuitively there should be an optimal value for restarting
probability.

One can verify that in this example the unique bounded solution to
(\ref{APZ2014DPEqn2}) is given by $V_1(x)=0$ for each $x\in[a,b]$,
\begin{eqnarray}\label{APZ2014Example111}
V_1(x)=\frac{1}{p}
\end{eqnarray}
for each $x>b$,
\begin{eqnarray}\label{APZ2014Example112a}
V_1(x) = \frac{1}{p} - \frac{1-p}{p} \left(1-e^{-\mu(b-a)}\right) e^{-\mu(a-x)p}
\end{eqnarray}
for each $x<a.$ In fact, this can be conveniently established
using the following probabilistic argument. Recall that $V_1(x)$
represents the expected total discounted time up to the hitting of
the set $H$ by the process $X$; see (\ref{eq:V1}). So for
(\ref{APZ2014Example111}) one merely notes that with the initial
state $x>b,$ $\eta^X_H=\infty$, where $\eta^X_H$ is defined by
(\ref{APZ2014NewNotation}). For (\ref{APZ2014Example112a}), one
can write for each $x<a$ that
\begin{eqnarray*}
V_1(x)&=&E_x\left[\sum_{t=0}^{\eta^X_H-1}
(1-p)^t\right]=\left.E_x\left[E_x\left[\sum_{t=0}^{\eta^X_H-1}
(1-p)^t\right |\eta_H^X\right]\right].
\end{eqnarray*}

Now the expected hitting time of the restarted process $\tilde{X}$
to the set $H=[a,b]$ is given by
\begin{eqnarray*}
V(x) = \left(\frac{1}{p} - \frac{1-p}{p}
\left(1-e^{-\mu(b-a)}\right)
e^{-\mu(a-x)p}\right)\sum_{t=0}^\infty \left(p\int_{E}
V_1(y)\nu(dy)\right)^t,
\end{eqnarray*}
with the initial state $x<a$, and by
\begin{eqnarray*}
V(x) = \frac{1}{p} \sum_{t=0}^\infty \left(p\int_{E}
V_1(y)\nu(dy)\right)^t,
\end{eqnarray*} with the initial state $x
> b$, recall (\ref{eq:explsol}). If the restart distribution $\nu$
is not concentrated on $(b,\infty)$, then $q(H)>0$, and by Theorem
\ref{thm:explsol}(b) we have
\begin{eqnarray*}
V(x) = \frac{\frac{1}{p} - \frac{1-p}{p}
\left(1-e^{-\mu(b-a)}\right) e^{-\mu(a-x)p}}{1-p \int_E V_1(y)
\nu(dy)},
\end{eqnarray*}
with the initial state $x<a$ and by
\begin{eqnarray*} V(x) =
\frac{1}{p \left(1-p \int_E V_1(y) \nu(dy) \right)},
\end{eqnarray*}
for each $x > b$. In particular, if the process restarts from a
single point $r < a$, the above expressions can be specified to
\begin{eqnarray*}
 V(x) = \frac{1 - (1-p)(1-e^{-\mu(b-a)}) e^{-\mu(a-x)p} }
{p(1-p)(1-e^{-\mu(b-a)}) e^{-\mu(a-r)p}},
\end{eqnarray*}
for the initial state $x<a$ and to
\begin{eqnarray*}
V(x) = \left(\frac{1}{1-e^{-\mu(b-a)}}\right)\left( \frac{1}{p
(1-p) e^{-\mu(a-r)p}}\right),
\end{eqnarray*}
for each $x>b$. For the latter case ($x>b$), by standard analysis
of derivatives, one can find the optimal value of the restart
probability minimizing the expected hitting time of the process
with restart in a closed form, as given by
\begin{eqnarray*}
p_{opt} = \frac{2}{2+\mu(a-r)+\sqrt{4+\mu^2(a-r)^2}}.
\end{eqnarray*}
Now we can make several observations: the first somewhat
interesting observation is that in the case when the initial state
is to the right of the interval $[a,b]$, the value of the optimal
restart probability does not depend on the length of the interval
but only on the average step size and on the restart position. The
second observation is that when $\mu(a-r)$ is small, i.e., when
either the average step size is large or the restart position is
close to $H$, the optimal restart probability is close to 1/2.
Thirdly, when $\mu(a-r)$ is large, the optimal restart probability
is small and reads
\begin{eqnarray*}
p_{opt} = \frac{1}{1+\mu(a-r)} + \mbox{o}\left(\frac{1}{\mu(a-r)}\right).
\end{eqnarray*}

\subsection{Random walk on the one dimensional lattice}

Let us consider a symmetric random walk on the one dimensional
lattice which aims to hit $H=\{0\}$ with restart at some node $r$.
Assume without loss of generality that the restart state $r$ is on
the positive half-line, i.e., $r>0$.

From Theorem~\ref{thm:explsol} we conclude that it is sufficient
to solve the following equations
\begin{eqnarray*}
&&V_1(k) = 1 + \frac{1-p}{2}[V_1(k-1)+V_1(k+1)], \quad k \neq 0, \\
&&V_1(0) = 0.
\end{eqnarray*}
Following the standard approach for solution of difference
equations, we obtain
\begin{eqnarray*}
V_1(k) = c  \alpha_1^k + \frac{1}{p},
\end{eqnarray*}
where $\alpha_1<1$ is the minimal solution to the characteristic
equation
\begin{eqnarray*}
\alpha = \frac{1-p}{2} [ 1 + \alpha^2 ],
\end{eqnarray*}
and the constant $c =-\frac{1}{p}$ comes from the condition
$V_1(0)=0$.  Consequently,
\begin{eqnarray*}
 V(k) = \frac{V_1(k)}{1-pV_1(r)} =
\frac{1-\alpha_1^k}{p \alpha_1^r}.
\end{eqnarray*}
An elegant analysis can be done for the limiting case when the
initial position $k$ goes large, and hence we now minimize
$\lim_{k\rightarrow \infty}V(k)=1/(p\alpha_1^r)$, or equivalently,
maximize $p\alpha_1^r$ with respect to $p\in(0,1).$ This leads to
the following equation for the optimal restart probability
\begin{equation}
\label{eq:poptex2}
\frac{p}{\alpha_1} \frac{d\alpha_1}{dp} = - \frac{1}{r}.
\end{equation}
Indeed, we note that
\begin{eqnarray*}
&&\lim_{p \to 0} \frac{p}{\alpha_1} \frac{d\alpha_1}{dp} =
0,~\lim_{p \to 1} \frac{p}{\alpha_1} \frac{d\alpha_1}{dp} =
-\infty,
\end{eqnarray*}
and
\begin{eqnarray*}
\frac{d
}{dp}\left(\frac{p}{\alpha_1} \frac{d\alpha_1}{dp}\right) =
-\frac{1}{\sqrt{1-(1-p)^2}}\frac{1-(1-p^2)(1-p)}{(1-p)^2(1-(1-p)^2)}
< 0.
\end{eqnarray*} Thus, the left hand side of
(\ref{eq:poptex2}) is a monotone function decreasing from zero to
minus infinity. Consequently, the unique solution of equation
(\ref{eq:poptex2}) is the global minimizer of $1/(p\alpha_1^r)$.
The equation (\ref{eq:poptex2}) can be transformed to the
polynomial equation
$$
\frac{1}{r^2}(1-p)^2(2-p)=p
$$
Consider the case of large $r$. This is a so-called case of singular perturbation,
as the small parameter $1/r^2$ is in front of the largest degree term,
\cite{AFH13}, \cite{B85}.
It is not difficult to see that for large values of $r$, the equation has one
real root that can be expanded as
\begin{equation}
\label{eq:poptseries}
p_{opt} = \frac{c_1}{r^2}+\frac{c_2}{r^4}+...
\end{equation}
and two complex roots that move to infinity as $r \to \infty$. By
substituting the series (\ref{eq:poptseries}) into the polynomial
equation, we can identify the terms $c_i, i=1,2,...$.
Thus, we obtain
\begin{eqnarray*} p_{opt} = \frac{2}{r^2} -
\frac{10}{r^4} + \mbox{o}(r^{-4}).
\end{eqnarray*}

\subsection{Application to network centrality}

\begin{figure}[ht]
\includegraphics[scale=0.5]{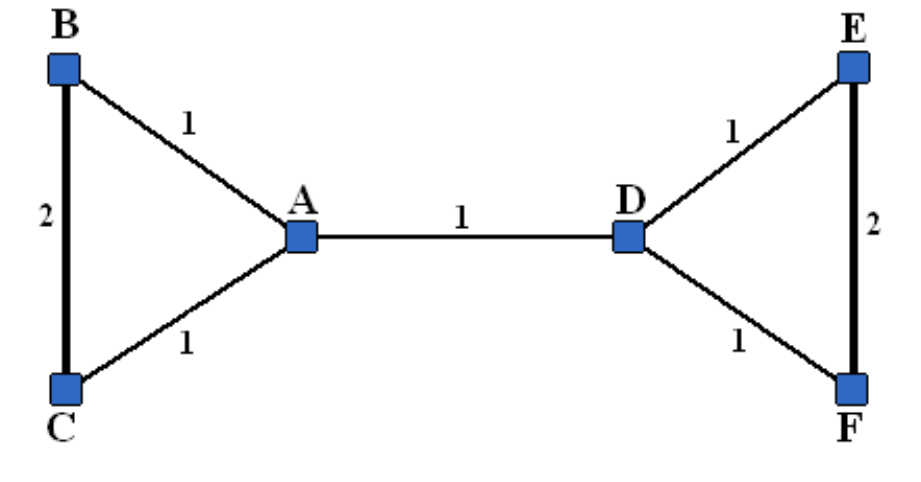}
\caption{Example of 6-node network with weighted edges.}
\label{fig:6nodesexample}
\end{figure}

One of the main tasks in network analysis is to determine which nodes are more ``central''
than the others. Node degree and PageRank \cite{BP98} are examples of widely used centrality
measures. We note that PageRank is a stationary distribution of the random walk with restart
and in our setting it is just measure $q$. Both node degree and PageRank are prone to
manipulation or so-called ``sybil attack''. To mitigate this problem, the authors of \cite{HS08,Liuetal16}
proposed hitting time based centrality measures. Here we show that the hitting time based
centrality can be more discerning. Let us first consider a simple 6-node network with weighted
edges, see Figure~\ref{fig:6nodesexample}. The weights are depicted near the edges.

In Table~\ref{tab:6nodesvalues} we give centrality values of the nodes with respect to node degree, PageRank
and Expected Hitting Times starting and restarting both from the uniform distribution.

\begin{table}[ht]
\caption{Centralities for 6-node network.}
\label{tab:6nodesvalues}
\begin{tabular}{|c|c|c|c|c|c|c|}
\hline
Nodes & A & B & C & D & E & F \\
\hline
Node degree & 3 & 3 & 3 & 3 & 3 & 3 \\
\hline
PageRank, $\forall p$ & 1/6 & 1/6	& 1/6	& 1/6	& 1/6	& 1/6 \\
\hline
Hitting time, $p=0.15$ & 6.28 & 8.28 & 8.28 & 6.28 & 8.28 & 8.28 \\
\hline
\end{tabular}
\end{table}
The last row in Table \ref{tab:6nodesvalues} lists down the expected hitting times to the target state A, B, and so on.
It is intuitively clear that nodes A and D are more central in this network. However, both node
degree and PageRank indicate equal importance for the nodes. In contrast, the hitting time
based centrality clearly indicates that nodes A and D are more central than the other nodes.

\begin{figure}[ht]
\includegraphics[scale=0.25]{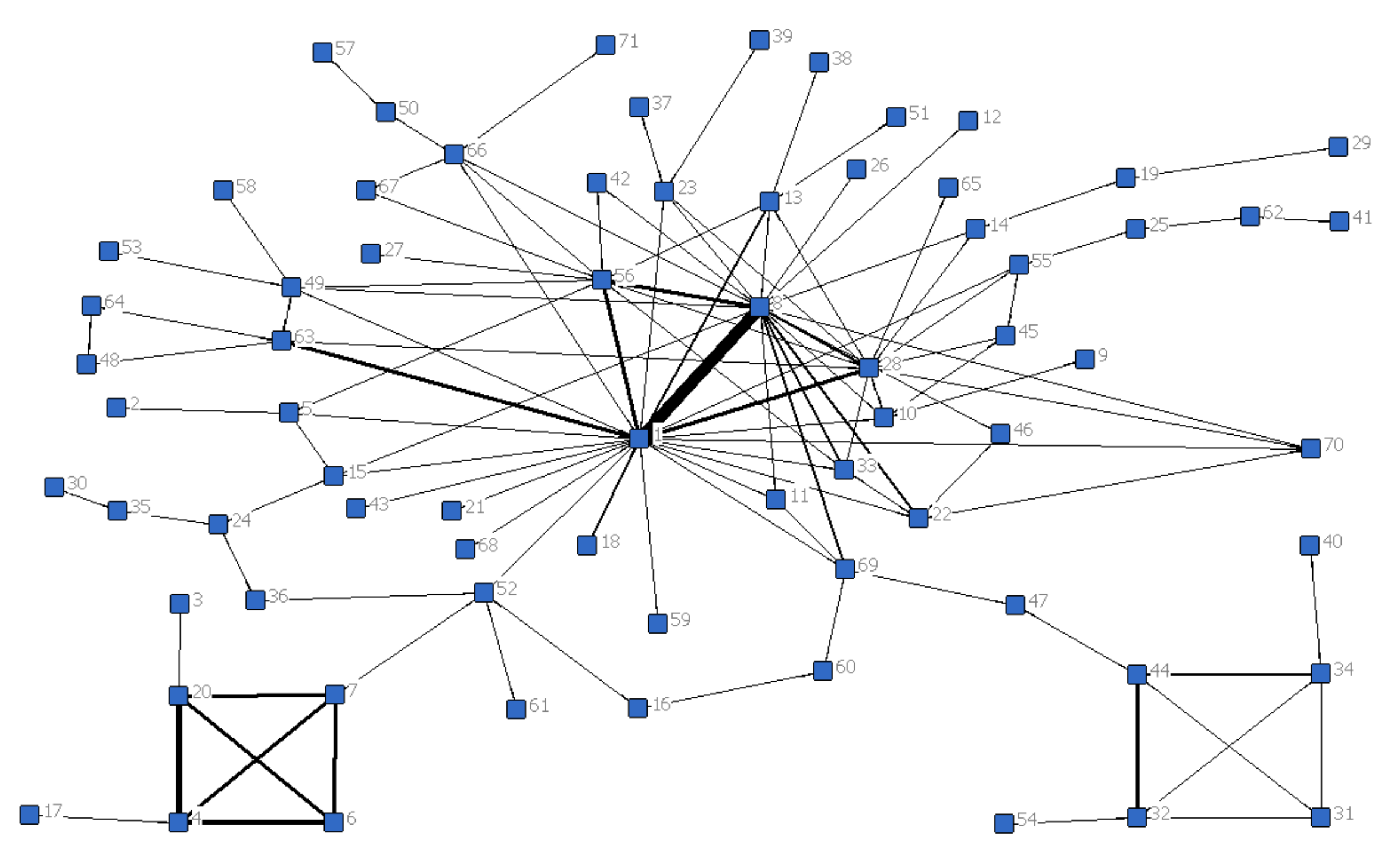}
\caption{Example of a social network.}
\label{fig:SocNetExample}
\end{figure}

Let us now consider an example of a real social network. The example was taken from online
social network VKontakte and represents a principal component of the interest group about
Game Theory \cite{AMT15}. The example has 71 nodes and 116 weighted edges,
see Figure~\ref{fig:SocNetExample} (taken from \cite{AMT15}).
The edge weight is equal to the number of common friends. Only the edges with a weight more
than two have been kept.

In Table~\ref{tab:top10lists} we provide top-10 lists of nodes according to node degree, PageRank
and the expected hitting time.

\begin{table}[ht]
\caption{Top-10 lists for the social network example.}
\label{tab:top10lists}
\begin{tabular}{|c|c|c|c|c|c|c|c|c|c|c|}
\hline
Node degree & 1 & 8 & 4 & 20 & 6 & 56 & 7 & 28 & 44 & 32 \\
\hline
PageRank, $p=0.15$ & 1 & 8 & 56 & 28 & 44 & 4 & 32 & 20 & 63 & 6 \\
\hline
Hitting time, $p=0.15$ & 1 & 8 & 56 & 28 & 63 & 22 & 13 & 33 & 69 & 4 \\
\hline
\end{tabular}
\end{table}

We observe that the top-10 list by PageRank has 9 nodes from the top-10 list by node degree.
The top-10 list by the expected hitting time has only 5 nodes from the top-10 list by node
degree. Note that nodes 22 and 13, which intuitively look quite central, are not in the top-10
list by PageRank.

\begin{figure}[ht]
\includegraphics[scale=0.6]{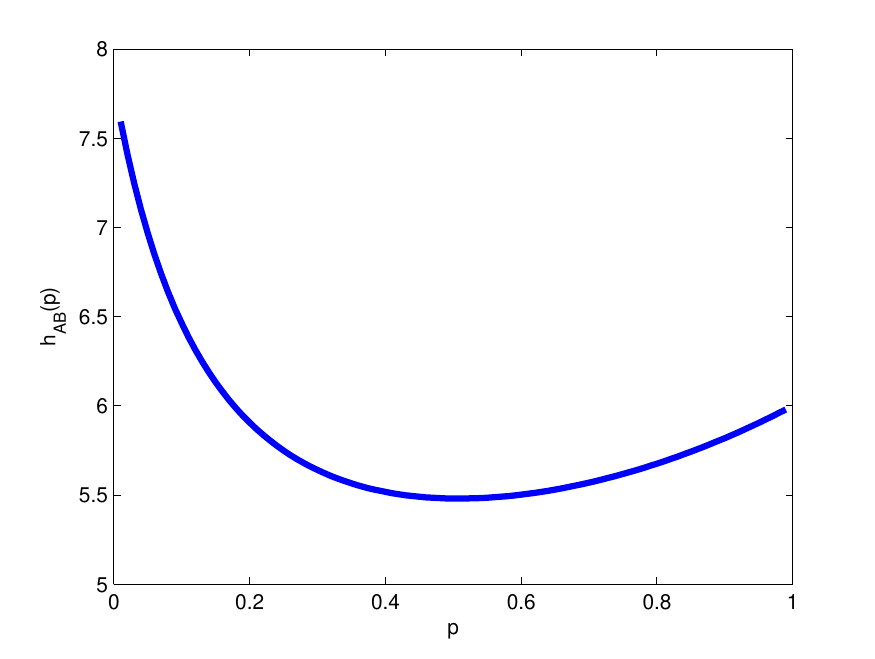}
\caption{The expected hitting time from A to B.}
\label{fig:hAB}
\end{figure}

Finally, we would like to show in Figure~\ref{fig:hAB} the expected hitting time from node
A to node B in the 6-node example as a function of the restart probability $p$. This function
has a minimum inside the interval $[0,1]$. We think it will be interesting to study the
minimization of the expected hitting time in the context of network community analysis.

\section{Conclusion}\label{APZ2014Conclusion}

In conclusion, in this paper we present a self-contained study of a discrete-time Markov process with restart.
Our primary interest is in the expected hitting time
of the process with restart to a target set. We obtained the formula of the
expected hitting time of the restarted process to a target set, and considered
the optimization problem of the expected hitting time with respect to the restart probability.
We illustrated our results with two examples in uncountable and countable state spaces and one
application to network centrality. In particular, we show that the network centrality based on
hitting times is more selective. Our general results may also have potential application to network
community analysis, which we intend to explore in the future.

\section*{Acknowledgements}
This work was partially supported by the European Commission
within the framework of the CONGAS project FP7-ICT-2011-8-317672. Y.Zhang's work
was carried out with a financial grant from the Research Fund for Coal and Steel of
the European Commission, within the INDUSE-2-SAFETY project (Grant No. RFSR-CT-2014-00025).

\tableofcontents

\end{document}